\documentclass[oribibl]{llncs}

\PassOptionsToPackage{hyphens}{url}
\usepackage[colorlinks = true,
  linkcolor = blue,
  citecolor = blue,
  urlcolor = blue]{hyperref}

\usepackage{url}
\usepackage{amsmath}
\usepackage{amssymb}
\usepackage{csquotes}
\usepackage{multirow}
\usepackage{amsmath}
\usepackage{graphicx}
\usepackage{tabularx}
\usepackage{booktabs}
\usepackage{threeparttable}

\begin{document}

\title{The Baby Steps of the European Union Vulnerability Database: An Empirical Inquiry}

\author{Jukka Ruohonen\orcidID{\scriptsize{0000-0001-5147-3084}} \\ \email{juk@mmmi.sdu.dk} \institute{University of Southern Denmark, S\o{}nderborg, Denmark}}

\maketitle

\begin{abstract}
A new European Union Vulnerability Database (EUVD) was introduced via a
legislative act in 2022. The paper examines empirically the meta-data content of
the new EUVD. According to the results, actively exploited vulnerabilities
archived to the EUVD have been rather severe, having had also high exploitation
prediction scores. In both respects they have also surpassed vulnerabilities
coordinated by European public authorities. Regarding the European authorities,
the Spanish public authority has been particularly active. With the exceptions
of Finland, Poland, and Slovakia, other authorities have not engaged thus
far. Also the involvement of the European Union's own cyber security agency has
been limited. These points notwithstanding, European coordination and archiving
to the EUVD exhibit a strong growth trend. With these results, the paper makes
an empirical contribution to the ongoing work for better understanding European
cyber security governance and practice.
\end{abstract}

\begin{keywords}
software vulnerabilities, vulnerability meta-data, vulnerability tracking,
exploitable vulnerabilities, actively exploited vulnerabilities, CSIRT, ENISA
\end{keywords}

\section{Introduction}

The EUVD was established by a requirement imposed by the second information and
network security (NIS2) directive.\footnote{~Directive (EU) 2022/2555.} It is
related to the broader new legal requirements for vulnerability coordination and
disclosure in Europe, including with respect to the CRA, that is, the Cyber
Resilience Act (CRA).\footnote{~Regulation (EU) 2024/2847.} The EUVD is
maintained by ENISA, that is, the European Union Agency for Cybersecurity.

The EUVD is noteworthy already because it is the first officially supported
vulnerability database covering the whole European Union (EU). Thus far and
largely still so in the Western countries but also elsewhere, the coordination
and archiving of vulnerabilities have occurred through institutions located in
the United States, including the non-profit MITRE corporation who is primarily
responsible for the Vulnerabilities and Exposures (CVEs) and the National
Vulnerability Database (NVD) maintained by the National Institute of Standards
and Technology (NIST) of the United States. Therefore, the EUVD also aligns with
the EU's aspirations for digital sovereignty and strategic
autonomy~\cite{Paulus25}. That said, many countries outside of Europe have long
maintained their own vulnerability databases. For instance, Japan has
one.\footnote{~\url{https://jvn.jp/en/}} China too has
one.\footnote{~\url{https://www.cnvd.org.cn/}} These and the Russia's
vulnerability database have occasionally prompted also media
attention~\cite{Leyden18}. Against these backdrops, the EUVD is hardly anything
new as such in the global arena. Regarding the EU itself, the EUVD is therefore
perhaps better framed against the new legislation, including the CRA and
NIS2~directive.

The new EU legislation provides the background and motivates the four research
questions (RQs) formulated in the subsequent Section~\ref{sec: background}. The
dataset used is elaborated in the following Section~\ref{sec: data}. The
empirical results are presented in Section~\ref{sec: results}. The final
Section~\ref{sec: conclusion and discussion} presents a conclusion and a brief
discussion.

\section{Background and Research Questions}\label{sec: background}

The NIS2 directive and the CRA impose various legal requirements for incident
notifications and vulnerability reporting, some of which are voluntary and some
of which are mandatory~\cite{Ruohonen25COSE}. Regarding vulnerabilities,
reporting of ``conventional'' vulnerabilities is voluntary, but reporting of
\textit{actively exploited vulnerabilities} (AEVs) is mandatory albeit with a
delaying option~\cite{Ruohonen25ACIG}. To ease reporting, a unified EU-level
platform has been established for AEVs and severe incidents~\cite{ENISA26a}. The
definition for AEVs emphasizes the existence of reliable evidence that
exploitation has already occurred.\footnote{~Article 3(42) in Regulation (EU)
2024/2847.} Therefore and in general, AEVs can be likened to the notion of
\textit{known exploited vulnerabilities} (KEVs) used in the United
States~\cite{CISA26}. These points motivate the first research question:
\begin{itemize}
\item{$\textmd{RQ}_1$}: How severe have the AEVs archived to the EUVD been?
\end{itemize}

The first $\textmd{RQ}_1$ is worth asking because a recent work found only a
weak statistical relationship between exploitation and
severity~\cite{Khoury25}. With this research question, the paper also aligns
with the rather extensive research on software vulnerabilities and their
severity. Within this severity-specific research, the topics examined range from
the relation between severity and exploits~\cite{Allodi14}, severity and actual
exploitation~\cite{Holm12, Khoury25}, severity and bug
bounties~\cite{Munaiah16}, and delays in severity
assessments~\cite{Ruohonen19ACI} to the validity of severity
metrics~\cite{Allodi14, Esposito23}, to note~a~few~examples.

The point about the validity of severity metrics characterizes also the
subsequent research question because also metrics about exploitation, including
probability of exploitation in the future in particular, have recently been
criticized~\cite{Massacci24}. While keeping this point in mind, it is still
relevant in an \textit{exploratory} manner to know how AEVs align with existing
exploitation prediction scores. By using the Exploit Prediction Scoring System
(EPSS), which tries to approximate the probability of exploitation in the wild
in the next 30 days, from the FIRST organization~\cite{Jacobs23}, the second
research question can be presented as follows:
\begin{itemize}
\item{$\textmd{RQ}_2$}: What have the EPSS scores of the AEVs archived to the EUVD been?
\end{itemize}

The NIS2 directive and the CRA strengthened also vulnerability coordination,
including vulnerability disclosure, within and across Europe. Both legislative
acts further strengthened ENISA's role as a EU-level coordination hub. At the
national level they strengthened the pan-European network of computer security
incident response teams (CSIRTs) through which vulnerabilities are primarily
coordinated in the EU via official, authoritative channels. Even though the
details are still being worked on, including with respect to unsolved issues
such as a lack of legal protections for reporters~\cite{Rampasek26}, the
adoption of the new legal requirements for vulnerability coordination seems to
be well underway~\cite{Neef25}. Given this background, it is generally
interesting to ask the following:
\begin{itemize}
\item{$\textmd{RQ}_3$}: Which European authoritative CSIRTs have been active in
  coordinating vulnerabilities archived to the EUVD and how active ENISA has
  been?
\end{itemize}

The final research question combines two of the previous RQs into:
\begin{itemize}
\item{$\textmd{RQ}_4$}: How severe have the vulnerabilities coordinated through
  the European CSIRT network and ENISA been, what their EPSS scores have been,
  and do the answers differ with respect to those for the earlier $\textmd{RQ}_1$ and $\textmd{RQ}_2$?
\end{itemize}

The four RQs are---out of necessity---exploratory because---to the best of the
author's knowledge---no prior empirical work exists about the EUVD. That said,
the forthcoming empirical examination of them carries practical relevance
because only little is known also about the workings of vulnerability
coordination in Europe. To this end, the paper fills a noteworthy gap in
existing research.

\section{Data}\label{sec: data}

The dataset, which is available online for replication
purposes~\cite{Ruohonen26DSICIST}, was collected on 15 December 2026 from the
EUVD's online website maintained by ENISA~\cite{ENISA26b}. Although the EUVD
archives many vulnerabilities, including supposedly via a synchronization with
the NVD and other sources, the analysis is restricted to AEVs and those
vulnerabilities coordinated via ENISA or the European CSIRT network. At the time
of the data collection, the database had $1,510$ AEVs and $1,291$
ENISA/CSIRT-coordinated vulnerabilities.

All of the vulnerabilities observed have CVE identifiers. This observation
further aligns AEVs to KEVs because the latter require CVEs to be
recognized~\cite{CISA26}. In addition, $325$ of the vulnerabilities observed
further have GitHub security advisory identifiers, suggesting that also
vulnerabilities about open source software are archived to the EUVD at least to
some extent.

Interestingly enough, furthermore, there are no overlaps between the AEVs and
the ENISA/CSIRT-coordinated vulnerabilities according to both the EUVD's
identifiers and CVEs. This observation signals that thus far the European CSIRT
network and ENISA have not been active in assessing whether the vulnerabilities
are under active exploitation. But this may change in the future, of course.

Regarding severity, the Common Vulnerability Scoring System (CVSS) from the
FIRST organization is used because it is also used in the EUVD. However, four
different versions are used in the database. Of these and with respect to
$\textmd{RQ}_1$ and $\textmd{RQ}_4$, four vulnerabilities archived with the
ancient CVSS v.~2.0 standard are excluded and those with CVSS v.~3.0 were
transformed into the 3.1 version. Then, the analysis is carried out separately
for the CVSS~3.1 and 4.0~versions.

\section{Results}\label{sec: results}

\begin{figure}[th!b]
\centering
\includegraphics[width=\linewidth, height=5cm]{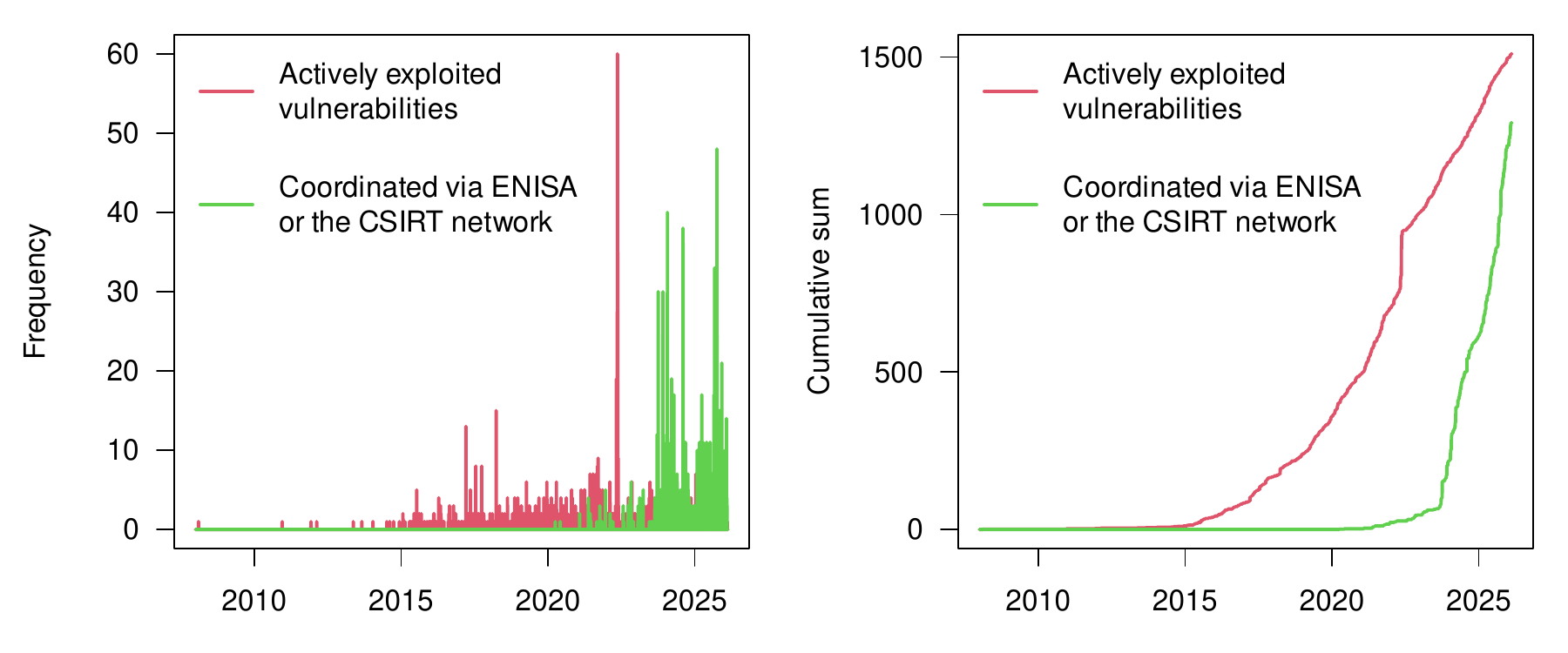}
\caption{Publication Dates of the Vulnerabilities}
\label{fig: dates}
\end{figure}

Descriptive statistics are sufficient for answering to the four research
questions. Before turning to these, it is useful to take a look at the
longitudinal evolution of the database. Thus, Fig.~\ref{fig: dates} shows the
publication dates of the vulnerabilities archived; these refer to those supplied
in the \texttt{datePublished} field of the EUVD's schema. As can be seen, the
archiving of AEVs started in around 2015, but regarding the EUVD itself, this
archiving is due to synchronization with other vulnerability databases. In
contrast, the archiving of vulnerabilities coordinated via ENISA or the European
CSIRT network started in late 2024. This observation aligns well with the
enactment of the NIS2 directive in late 2022. The growth rate of these
vulnerabilities has been rapid ever since. Given the time of writing in
mid-February 2026, the amount of these vulnerabilities archived to the EUVD have
almost reached the amount of actively exploited vulnerabilities.

\begin{figure}[th!b]
\centering
\includegraphics[width=\linewidth, height=5cm]{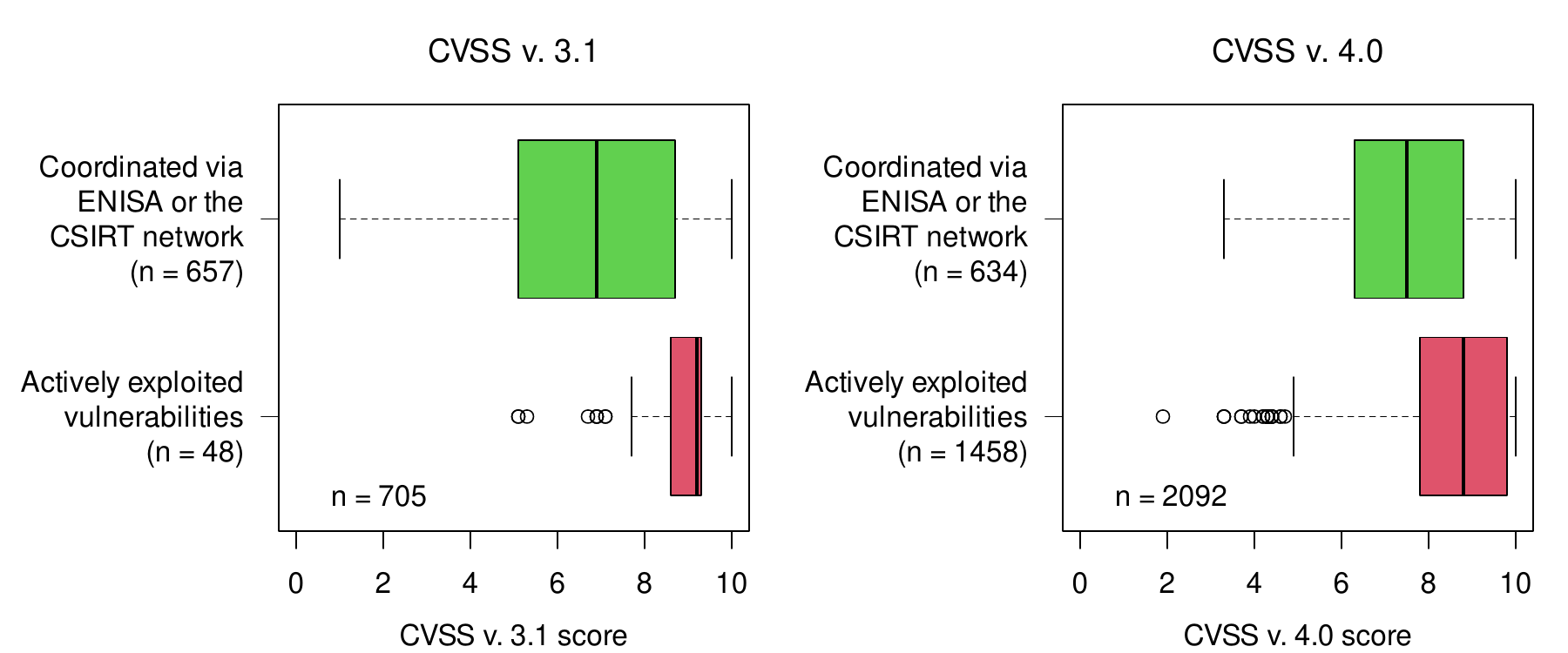}
\caption{Severity of the Vulnerabilities}
\label{fig: severity}
\end{figure}

\begin{figure}[p!]
\centering
\includegraphics[width=\linewidth, height=7cm]{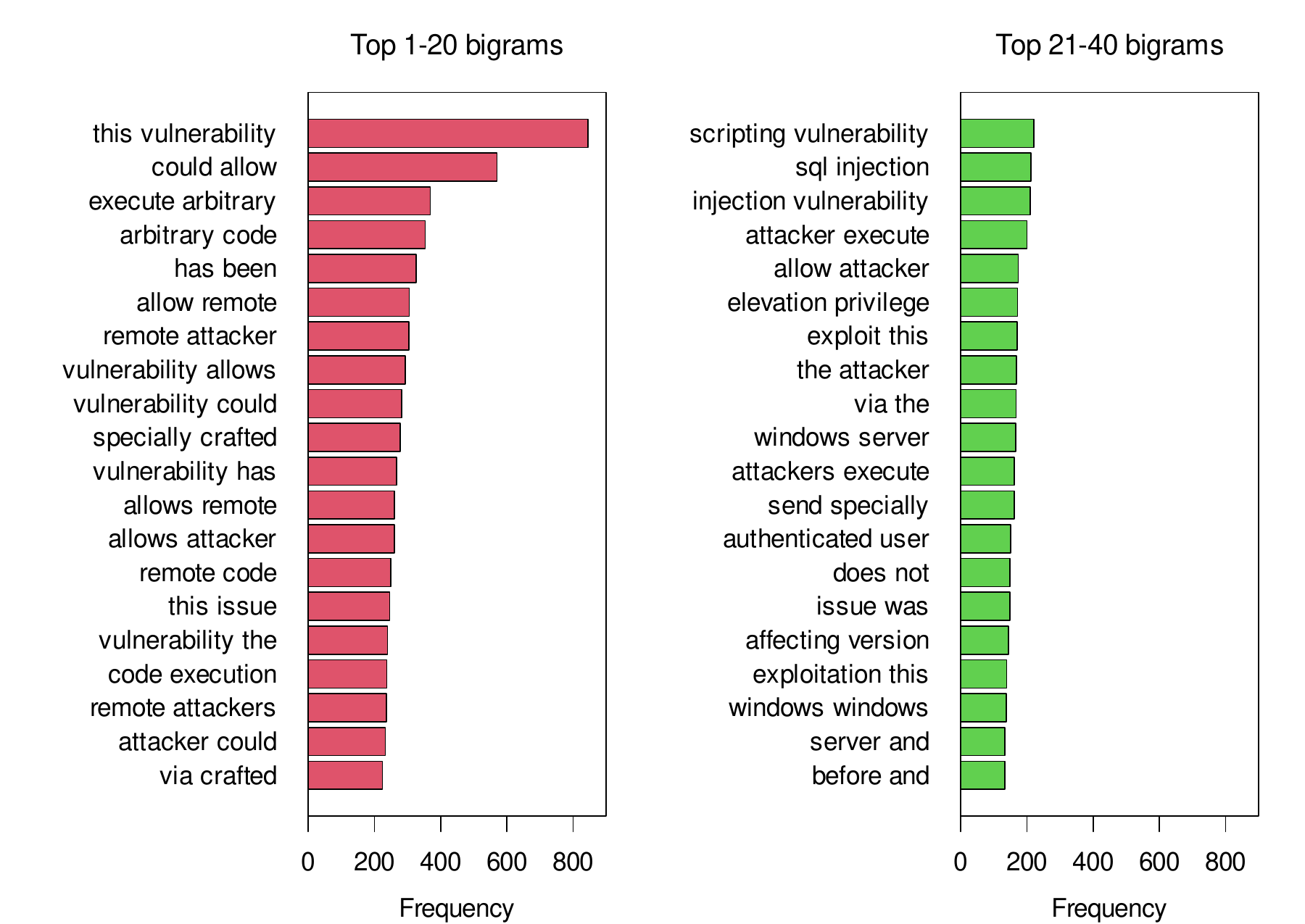}
\caption{Most Frequent Bigrams in the Vulnerabilities' Descriptions}
\label{fig: bigrams}
%
\vspace{30pt}
%
\centering
\includegraphics[width=\linewidth, height=3.5cm]{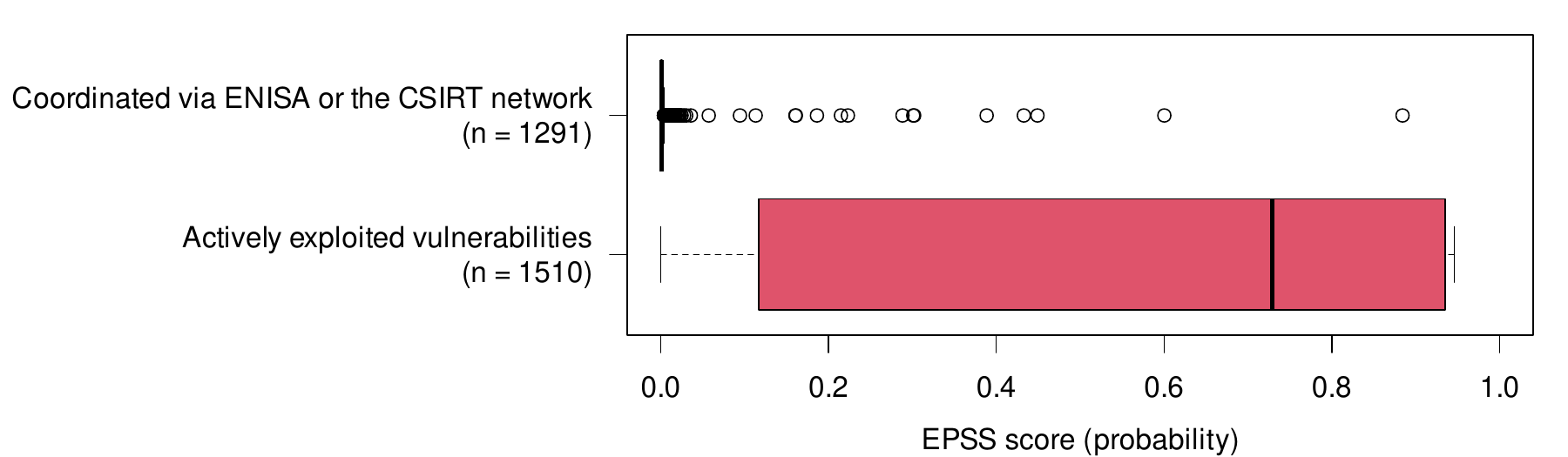}
\caption{EPSS Scores of the Vulnerabilities}
\label{fig: epss}
%
\vspace{30pt}
%
\centering
\includegraphics[width=\linewidth, height=3.5cm]{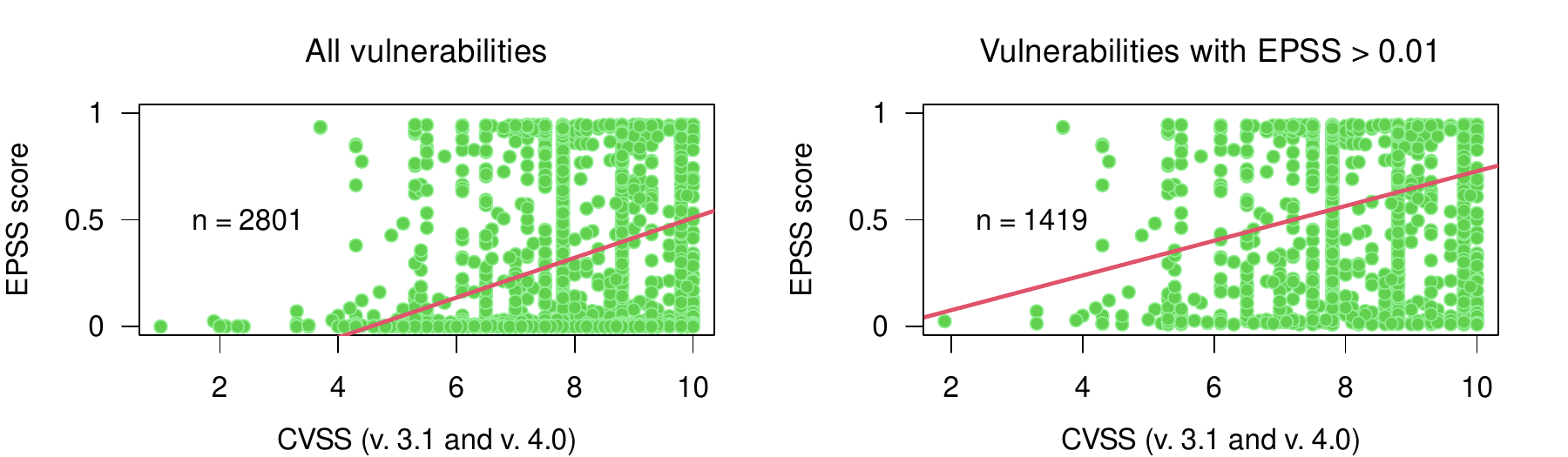}
\caption{Correlations Between CVSS and EPSS Scores}
\label{fig: cor}
\end{figure}

\begin{figure}[th!b]
\centering
\includegraphics[width=\linewidth, height=7cm]{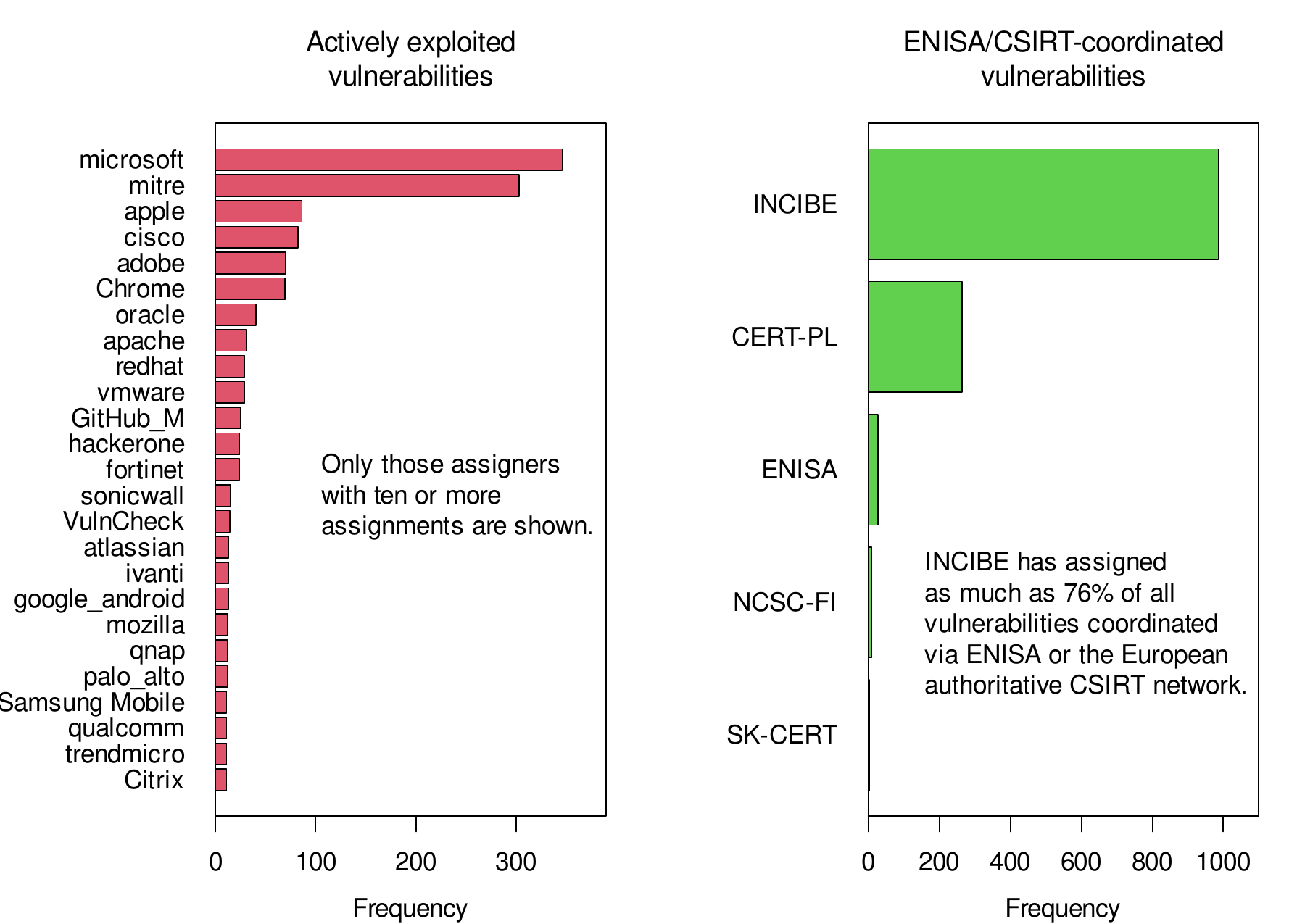}
\caption{Assigners of the Vulnerabilities}
\label{fig: assigners}
\end{figure}

Turning to $\textmd{RQ}_1$, Fig.~\ref{fig: severity} allows to conclude that the
AEVs archived to the EUVD have been rather severe on average; the medians are
well-beyond eight. Although it is difficult---if not dubious---to attach
adjectives to CVSS scores, the wording about ``rather severe'' can be justified
by taking a brief look at the literature. Thus, many vulnerabilities affecting
the Linux kernel have received CVSS v.~3.0 scores less than
eight~\cite{Azizi25}, whereas many web vulnerabilities have hardly surpassed a
CVSS v.~2.0 value four sometimes considered as a threshold for the importance
and urgency of patching~\cite{Ruohonen18IWESEP}. The most frequent bigrams shown
in Fig.~\ref{fig: bigrams} further shed some tentative light on the severity;
remote code execution vulnerabilities seems to rank high according to the
frequencies shown.\footnote{~The bigrams were constructed from the
\texttt{description} field in the EUVD's schema. White space was used for
tokenization, and tokens with lengths less than three characters or more than
twenty characters were excluded. No other processing was~done.}

It can be also concluded from Fig.~\ref{fig: severity} that AEVs have been more
severe than the ENISA/CSIRT-coordinated vulnerabilities. The same conclusion can
be drawn from Fig.~\ref{fig: epss} that shows the EPSS scores; AEVs have much
higher scores. The mean and median EPSS scores for AEVs are $56$ and $73$,
respectively. Despite the criticism on exploitation metrics, these numbers allow
to tentatively argue that the EPSS framework may well have its merits. It may
also be useful for empirical research purposes. For instance, there is a
positive correlation between the CVSS and EPSS scores, as seen from
Fig.~\ref{fig: cor}. By recalling the wording about reliable evidence of
exploitation used in the EU legislation, further research is still required for
determining whether the EPSS framework can empirically differentiate between
AEVs and non-AEVs, and whether and how well EPSS scores might predict a
probability of a vulnerability ending up as an AEV.

The answer to $\textmd{RQ}_3$ can be deduced from the left-hand side plot in
Fig.~\ref{fig: assigners}. The Spanish National Cybersecurity Institute (INCIBE
via the Spanish name) has been particularly active; it has assigned a clear
majority of vulnerabilities coordinated by the European CSIRT network. Poland's
CSIRT (CERT-PL) takes the second place. Also Finland (NCSC-FI) and Slovakia
(SK-CERT) have coordinated a few vulnerabilities. Although the new EU
legislation established also some vulnerability coordination and disclosure
responsibilities for ENISA~\cite{Ruohonen25ACIG}, this EU-level agency has not
been active at the task thus far. In fact, ENISA has coordinated only about
2.1\% of all vulnerabilities archived to the EUVD.

Finally, regarding AEVs, the left-hand side plot in Fig.~\ref{fig: assigners}
allows to conclude that many of the assignments have come directly either from
large vendors or through the MITRE's global coordination setup. Of the vendors,
Microsoft and to a lesser extent Apple, Cisco, and Adobe have been particularly
active in assigning AEVs. That said, documentation is lacking about how
assignments for actively exploited vulnerabilities are done in practice
particularly by vendors themselves. This topic warrants further research,
including also because of the strict deadlines and potential sanctions imposed
by the EU's new legislation~\cite{Ruohonen25COSE}.

\section{Conclusion and Discussion}\label{sec: conclusion and discussion}

The paper examined empirically the EUVD's meta-data with four research questions
in mind. With respect to $\textmd{RQ}_1$ and $\textmd{RQ}_2$, the answers are
clear enough: actively exploited vulnerabilities archived to the EUVD have been
rather severe, having had also high EPSS scores for \textit{a~priori} predicting
a probability of exploitation. These results seem logical. Actively exploited
vulnerabilities have also been more severe than the ENISA/CSIRT-coordinated
vulnerabilities, and they have further had much higher EPSS
scores~($\textmd{RQ}_4$). Regarding $\textmd{RQ}_3$ and the European public
authorities involved in the CSIRT network, the Spanish INCIBE has been
particularly active. In fact, it has assigned more vulnerabilities than all
other public authorities combined. To a lesser extent, the Polish public
authority has been active too. The authorities in Finland and Slovakia have
assigned and coordinated a few vulnerabilities too. Thus far, ENISA has not
taken a particularly active role despite new legal provisions provided for
it. That said, the new unified reporting platform~\cite{ENISA26a} may change the
situation in the future.

What about further research? Three research paths seem promising. The first is
about studying how future practice will align with the new EU legislation. The
perhaps most obvious but at the same time the perhaps most important question is
whether the EUVD will improve the European cyber security situation. At the
moment, many organizations covered by the NIS2 directive still have notable
problems in patching of vulnerabilities~\cite{ENISA25}. Yet, there is more in this path.

As was noted, the details about AEV assignments, starting from the notion of
reliable evidence, remain unclear at the time of writing. The CRA is mostly
silent about how to determine active exploitation, including perhaps with a
threshold for seriousness. Yet, it is clear that security and vulnerability
research, security testing, and security engineering in general are out of
scope; vulnerabilities ``that are discovered with no malicious intent for
purposes of good faith testing, investigation, correction or disclosure to
promote the security or safety of the system owner and its users should not be
subject to mandatory notification''.\footnote{~Recital 68 in Regulation (EU)
2024/2847.} While the availability of exploits is a well-studied
topic~\cite{Householder20}, less is known about exploitation, although some
notable works existing also in this regard~\cite{Bilge12}. As various data
sources about AEVs and KEVs have been used in recent research~\cite{Khoury25}, a
better understanding is needed on how public authorities and vendors determine
active exploitation and how robust the data sources are for research purposes.

The second research path is related: the EUVD should be validated, including
with respect to its meta-data's robustness and timeliness. Given existing
results indicating different bugs and problems in other vulnerability databases,
including the NVD~\cite{Anwar22, Massacci13}, a motivation for validation
research would be easy to establish. This research path might provide also
practical relevance because knowing about potential problems early on can save
time and resources later on. In other words, switching costs, including with
respect to different schemas or versions of standards, are a known issue also
with vulnerability databases~\cite{Ruohonen19ACI}.

Last but not least, the EUVD provides valuable insights into the future of cyber
security governance and coordination in the European Union. The NIS2 directive
and the CRA are noteworthy in this regard because they operate with a rather
complex multi-level governance framework~\cite{Ruohonen25COSE, Teichmann25}. As
was demonstrated, it is possible to observe via the database the engagement and
activity of different European public authorities in vulnerability
coordination. To this end, further research is required to deduce whether other
European public authorities will engage more in the future; whether there will
be further concentration of activity to some member states; and generally
whether convergence will occur. While there is existing research on the
coherence and convergence of European cyber security policies~\cite{Carrapico17,
  Ruohonen24I3E}, less is known about coherence and convergence in practice,
including with respect to the day-to-day cyber security~work.

\bibliographystyle{splncs03}

\end{document}